\documentclass[aps,prl,twocolumn,groupedaddress,showpacs]{revtex4}
\usepackage{graphicx}
\usepackage{dcolumn}
\usepackage{bm}

\def\bc{\begin{center}}
\def\ec{\end{center}}
\def\beq{\begin{equation}}
\def\eeq{\end{equation}}

\begin{document}


\title{Suppression of magnetotransport in strongly disordered graphene}

\author{K. Ziegler}%
\email{Klaus.Ziegler@Physik.Uni-Augsburg.de}
\affiliation{
Institut f\"ur Physik, Universit\"at Augsburg, D-86135 Augsburg, Germany
}

\date{\today}

\begin{abstract}
A tight-binding model with randomly fluctuating atomic positions is studied to discuss
the effect of strong disorder in graphene. We employ a strong-disorder expansion for the
transport quantities and find a diffusive behavior, where the conductivity is decreasing 
with increasing disorder.  For sufficiently strong disorder the magnetic field drops out of
the diffusion coefficient and the conductivity. This signals a strong suppression
of magnetotransport effects, a result which
is consistent with recent experimental observations by Morozov {\it et al.}
\end{abstract}

\pacs{81.05.Uw, 71.55.Ak, 72.10.Bg, 73.20.Jc}
\maketitle


{\it Introduction:}
Experimental studies on graphene revealed
remarkable transport properties of this material \cite{novoselov05,zhang05,geim07}.
There is a robust minimal conductivity in the absence of an external magnetic field
and a quantum Hall effect in the presence of a strong magnetic field 
\cite{novoselov05,zhang05}. Morozov {\it et al.} found that also for weak magnetic fields 
unusual transport properties exist: The peak of the magnetoresistance at zero field, the
hallmark of weak localization in two-dimensional systems \cite{altshuler}, is strongly
suppressed in graphene \cite{geim07,morozov06,cho07}. This suppression was attributed to
disorder caused by ripples in the graphene sheet. When these ripples are removed, the
quasiparticle mobility is significantly increased and
a normal magnetoresistance peak appears \cite{morozov06}.
The peak was also observed in multilayers of graphene, where ripples are less developed
because of the higher rigidity of the material.

In this letter we study the transport properties of graphene in terms of a tight-binding
model for quasiparticles on a honeycomb lattice in the presence of a
homogeneous magnetic field. Strong disorder is introduced by randomly fluctuating hopping rates. 
The study is restricted to a system without a gate (i.e. the system is at the Dirac point), 
where the model has a chiral symmetry.

The usual approximation by Dirac fermions is not applicable to strong randomness. 
Nevertheless, we find a qualitative agreement of the transport behavior between the 
tight-binding model and Dirac fermions with random vector potential \cite{ludwig94}
or random mass \cite{ziegler98}. This indicates that
the sublattice structure, and not the chiral symmetry, is responsible for the suppression
of magnetoresistance peak. 

{\it Tight-binding model on a honeycomb lattice:}
Hopping of quasiparticles on a honeycomb lattice is defined by the Hamiltonian
\beq
{\cal H}=-\sum_{R,R'}h_{R,R'}c_R^\dagger c_{R'} + h.c. 
\label{ham0000}
\eeq
with quasiparticle creation (annihilation) operators $c_R^\dagger$
($c_{R}$). The hopping rate $h_{R,R'}$ between lattice sites $R$ and $R'$ may fluctuate
from bond to bond around the average value $t$ due to ripples in the graphene sheet.

A non-orthogonal basis $\{ a_1,a_2\}$ (cf. Fig. 1) is used to express the hopping rates 
in terms of the sublattice representation by writing $R= r$ ($=r+b_1$) if $R$ on A (B).
Now $r$ can be expressed in the basis $\{ a_1,a_2\}$, and $h_{R,R'}$
in Eq. (\ref{ham0000}) is replaced by the matrix
\beq
H=\pmatrix{
0 & d+\Delta \cr
d^T+\Delta^T & 0 \cr
} ,
\hskip0.5cm
d_{r,r'}=t\sum_{j=1,2,3}\delta_{r',r-c_j} 
\label{hamilton2}
\eeq
for $c_1=0$, $c_2=a_2$, and $c_3=a_1+a_2$. 
$\Delta_{r,r'}=\sum_{j=1,2,3}t'_{r',r}\delta_{r',r-c_j}$ are random fluctuations
of the hopping term
with $\langle t'_{r',r}\rangle=0$. We set $t=1$ and keep in mind that $t'$ and all other
energies of the model are measured now in units of the hopping rate $t$.

Although curvature can have
significant effects \cite{cortijo06,dejuan07,castroneto07b}, it will be ignored here by assuming that the
lattice is flat on length scales that are relevant for transport properties. Therefore, 
only fluctuations of the hopping rates between neighboring sites on the honeycomb 
lattice are considered.

The quasiparticle Hamiltonian $H$ consists of a translational-invariant part 
$H_0\equiv \langle H\rangle$ and a term that describes the random fluctuations. 
For the latter we assume that on large scales it can be approximated by on-site
terms, provided there is only short-range disorder. Such a choice leads to 
a random vector potential, previously proposed for ripples in graphene 
by Morozov {\it et al.} \cite{morozov06} and derived for changes in the hopping
due to hybridization between different orbitals in curved graphene sheets and
due to elastic strain \cite{castroneto07b}: 
\beq
H=H_0+v_1\sigma_1 +v_2\sigma_2 
\label{hamilton}
\eeq
with Pauli matrices $\sigma_j$.
We introduce an uncorrelated Gaussian distribution for $v$ with mean zero 
and variance $g$ and write $v_1=v\cos\alpha$, $v_2=v\sin\alpha$ with a tunable
parameter $\alpha$.   
Similar models, based on Dirac fermions with a random vector potential, have been studied
in the literature \cite{ludwig94,katsnelson07,castroneto07}. The seminal
work by Ludwig {\it et al.} 
states that a random vector potential does not renormalize the variance $g$, and that
(using a bosonization approach in replica space) the conductivity is not affected by this
type of disorder \cite{ludwig94}. The density of states, on the other hand, 
has a power law near the
Dirac point with exponent $(\pi-g)/(\pi+g)$. This indicates that for strong disorder
($g>\pi$) the density of states diverges at the Dirac point, whereas it vanishes for weak
disorder ($g<\pi$). Physical implications of this behavior can be discussed in terms
of the Einstein relation, where the conductivity $\sigma$ is given as the product of density 
of states $\rho$ and diffusion coefficient $D$ as $\sigma\propto \rho D$. Then the nonzero
conductivity at the Dirac point (minimal conductivity) would imply an infinite $D$ for 
$g<\pi$ and $D=0$ for $g>\pi$. In other words, these findings describe a transition at 
$g=\pi$ from ballistic transport directly to localization, without an intermediate 
diffusive regime. 

For small $g$ the exponent of the density of states can also be evaluated in perturbation
theory as $1-g/\pi$. 
However, in a recent work on the density of states a disorder created energy scale 
$\exp(-\pi/g)/g$ was detected, below which perturbation theory breaks down \cite{dora08}. 
In other
words, the assumption of a power law is invalid at the Dirac point and in its vicinity.
The energy scale is similar to the Kondo scale of the Kondo effect \cite{hewson}, 
and was also found independently for the conductivity by Auslender and Katsnelson
\cite{auslender07}. It implies a nonzero density of states at the Dirac point.
A similar result was found for topological disorder \cite{cortijo06}. In the following, 
it shall be discussed that the nonzero density of states is associated with spontaneous
breaking of a symmetry which leads to diffusion of quasiparticles.
\begin{figure}
\centering
\includegraphics[width=0.1\textwidth]{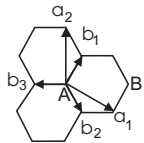}
\caption{Lattice vectors on the honeycomb lattice. The lattice is divided into
sublattice A and B, where nearest neighbors of A belong to sublattice B.}
\label{plotcond}
\end{figure}

{\it Magnetic field:}
A magnetic field $B$ perpendicular to the graphene plane 
enters the Hamiltonian through a Peierls phase, i.e. the hopping terms in $H_0$ are replaced by
\beq
d_{rr'}(B)=\sum_je^{i\phi_{r,j}}\delta_{r',r-c_j}, 
\hskip0.5cm
\phi_{r,j}=\int_{r}^{r+b_j} A\cdot db_j \ ,
\label{peierls0}
\eeq
where $A$ is the vector potential of the magnetic field $B$. A homogeneous magnetic field
creates additional terms to $H$ which are periodic in one spatial direction
and proportional to $\sigma_1$ and $\sigma_2$, respectively.

{\it Symmetries:}
$H$ is invariant under the transformation 
$H\to e^{i\varphi\sigma_3}He^{i\varphi\sigma_3}$. This is related to the
chiral symmetry discussed, for instance, in Refs. \cite{altland97,garcia06}. Here we
are not so much interested in the symmetry of the Hamiltonian
but in that of the two-particle Green's function (2PGF) at the Dirac point
\beq
K(r,r';\epsilon)
=\langle {\rm Tr}_2[G(r,r';i\epsilon)G(r',r;-i\epsilon)]\rangle \ .
\label{2pgf}
\eeq
${\rm Tr}_2$ is the trace with respect to the spinor structure, 
$G(i\epsilon)\equiv (H+i\epsilon)^{-1}$ is the one-particle
Green's function, and $\langle ...\rangle$ is the 
average with respect to the random term of the Hamiltonian. 
The second Green's function in Eq. (\ref{2pgf})
can be replaced by $-\sigma_3G(r',r;i\epsilon)\sigma_3$,
since it obeys the relation $G(-i\epsilon)=-\sigma_1 G(i\epsilon)\sigma_1$.
Within a linear-response approach $K(r,r';\epsilon)$ gives the transport 
properties: The DC conductivity at zero temperature is  
\cite{ludwig94,ziegler06,katsnelson05,gusynin05,falkovsky06,cserti06}
\beq
\sigma_{jj}
\propto \epsilon^2\sum_r r_j^2K(r,0;\epsilon)|_{\epsilon=0} \ .
\label{kubo0}
\eeq
For localized states the 2PGF decays exponentially on the localization length for which
this expression gives a vanishing DC conductivity.

To analyze the 2PGF in detail, we follow Ref. \cite{ziegler97} and 
write it, before averaging, as a Gaussian functional integral with
two independent Gaussian fields, a boson (complex) field $\chi_{rk}$ and a fermion 
(Grassmann) field $\Psi_{rk}$ ($k=1,2$):
\[
G_{jj'}(r,r';i\epsilon)G_{k'k}(r',r;i\epsilon)
\]
\beq
=\int\chi_{r'j'}{\bar \chi}_{rj} \Psi_{rk} {\bar \Psi}_{r'k'}\exp(-S_0)
{\cal D}\Psi {\cal D}\chi \ .
\label{finta}
\eeq
$S_0$ is a quadratic form of the four-component field 
$\phi_r=(\chi_{r1},\chi_{r1},\Psi_{r2},\Psi_{r2})$
\begin{equation}
S_0=
-i\sum_{r,r'}
\phi_r\cdot({\bf H}+i\epsilon)_{r,r'}{\bar\phi}_{r'} \ \ (\epsilon>0)\ .
\label{ssa0}
\end{equation}
The use of the mixed field $\phi_r$ avoids an extra normalization factor for the integral. 
(The replica trick, used in Ref. \cite{cortijo06}, is an alternative to avoid this
extra normalization factor.)
The extended Hamiltonian ${\bf H}=diag(H,H)$ of $S_0$ acts 
in the boson and in the fermion sector separately. It is invariant under 
the transformation 
\beq
{\bf H}\to {\bf U}{\bf H}
{\bf U}={\bf H}
{\ \ \ \ }{\rm with}\ \ {\bf U} 
=\exp\pmatrix{
0 & \psi\sigma_3\cr
{\bar\psi}\sigma_3 & 0 \cr
}
\label{supers}
\eeq
for Grassmann variables $\psi$ and ${\bar\psi}$, whereas $i\epsilon$ in Eq. (\ref{ssa0}) 
causes a symmetry breaking. This symmetry
is central for the transport properties, since it determines the large-scale properties 
of the 2PGF by creating a massless mode. The latter is a consequence of a possible
spontaneous symmetry breaking at the Dirac point $\epsilon\to0$ \cite{ziegler98}. 
This indicates that transport
at the Dirac point is {\it qualitatively} different from transport away from the
Dirac point, where this symmetry is explicitly broken by the Fermi energy. 
An important question in this context is how the symmetry affects average quantities,
and whether or not spontaneous symmetry breaking occurs. The signature of the latter is
a nonzero density of states at the Dirac point \cite{ziegler96}.
(The method of Ref. \cite{ziegler96} can be applied to the present model to prove 
that the average density of states is nonzero.)

Averaging the 2PGF over the Gaussian distribution of $v_r$ means
replacing $\exp(-S_0)$ by $\langle \exp(-S_0)\rangle$ on the right-hand side of 
Eq. (\ref{finta}). The average quantity can be written again as an exponential function
$\langle \exp(-S_0)\rangle=\exp(-S_1)$, where the new function $S_1$ contains also
quartic terms of the field $\phi$, a consequence of the Gaussian distribution.  
Then it is convenient to transform the integration variables as 
\beq
\pmatrix{\chi_r{\bar\chi}_r&\chi_r{\bar\Psi}_r\cr
\Psi_r{\bar\chi}_r&\Psi_r{\bar\Psi}_r\cr
}\rightarrow
{\bf Q}_r=\pmatrix{
Q_r & \Theta_r\cr
{\bar\Theta}_r & -iP_r\cr
} \ ,
\eeq
where $Q_{r}$, $P_{r}$ are symmetric $2\times2$ matrices, and $\Theta_{r}$, ${\bar\Theta}_{r}$
are $2\times2$ matrices whose elements are independent Grassmann variables. 
The average 2PGF now reads as a correlation function in the new field ${\bf Q}_r$: 
\beq
K(r,r';\epsilon)
=-\frac{1}{g^2}\int {\rm Tr}_2(\Theta_{r'}\gamma)
{\rm Tr}_2({\bar\Theta}_{r}\gamma)\exp(-S_2){\cal D}[{\bf Q}]
\label{green22}
\eeq
with
\beq
S_2=\sum_{r,r'}\frac{1}{g}{\rm Trg}({\bf Q}_{r}^2) 
-\ln[{\rm detg}[H_0+i\epsilon -2\gamma{\bf Q}]] \ .
\label{action2}
\eeq
${\rm Trg}$ is the graded trace, ${\rm detg}$ the graded determinant \cite{ziegler97},
and $\gamma$ is a $4\times4$ matrix, consisting of the linear combination
$\gamma_1\cos\alpha+\gamma_2\sin\alpha$, where $\gamma_j=diag(\sigma_j,\sigma_j)$. 
The integration over ${\bf Q}$ can be performed in a saddle-point (SP) approximation,
where the SP satisfies the extremal condition $\delta_{\bf Q} S_2=0$,
which reads explicitly
\beq
{\bf Q}_r^{SP}=2g[H_{0}+i\epsilon -2\gamma{\bf Q}^{SP}]^{-1}_{rr}\gamma \ .
\label{spe0}
\eeq
A consequence of the symmetry (\ref{supers}) is that with ${\bf Q}^{SP}$ also
${\bf U}{\bf Q}^{SP}{\bf U}^\dagger$ is a solution of the SP equation at the Dirac point 
$\epsilon=0$.  

The right-hand side of Eq. (\ref{spe0}) is translational invariant
in the absence as well as in the presence of a homogeneous magnetic field.
This is a consequence of the fact that the diagonal Green's function $G_{rr}$ describes
closed loops of quasiparticles. These loops depend only on the flux which they enclose. 
The translational-invariant $G_{rr}$ has a constant SP solution
$
{\bf Q}^{SP}=-i(\eta/2)\gamma
$ which satisfies
\beq
\eta=4ig(H_0+i\epsilon+i\eta)^{-1}_{rr} \ .
\label{spe1}
\eeq
$\eta$ can be interpreted as a self energy of the average one-particle Green's
function $(H_0+i\epsilon+i\eta)^{-1}$ \cite{peres06}. 
Then the average density of states at the Dirac point is $\rho=\eta/g$.
A SP solution $\eta>0$ exists even in the limit $\epsilon\to0$, which reflects 
spontaneous symmetry breaking. $\eta$ increases with $g$ and reaches for large $g$
the asymptotic regime with $\eta\sim 2\sqrt{g}$, where the density of states
decays like $\rho\sim g^{-1/2}$.

The invariance of the SP equation requires the integration over all SPs and
their vicinities. This leads to an SP manifold, generated by the symmetry transformation
in form of the nonlinear field
\beq
{\bf Q}_{r}'=-i{\eta\over2} {\bf U}^\dagger _r\gamma{\bf U}_{r} 
=-i{\eta\over2} \gamma{\bf U}_r^2 \ .
\label{spmf}
\eeq
${\bf U}_r$ is the matrix ${\bf U}$ of Eq. (\ref{supers}) with space-dependent 
Grassmann fields $\psi_r$ and ${\bar\psi}_r$. 
The integration with respect to the chiral symmetry is not taken
into account here because the corresponding fields are perpendicular 
to the Grassmann fields in leading order.
Thus they do not contribute to the average 2PGF. 

Replacing ${\bf Q}_r$ by ${\bf Q}_r'$ in $S_2$ provides two major 
simplifications: (1) the first term in expression (\ref{action2}) vanishes,
since ${\bf Q}_r'^2$ is proportional to the $4\times4$ unit matrix. (2) the
second term becomes
\beq
-\ln {\rm detg}({\bf U}^{-1}({\bf H}_0+i\epsilon){\bf U}^{-1}+i\eta) ,
\label{det1}
\eeq
since ${\rm detg}{\bf U}=1$. This expression can be expanded either in powers of
$\eta^{-1}$ or in powers of $\eta$. The problem of the latter case is that at low
energies, which are relevant at the Dirac point, the expansion terms 
$({\bf H}_0+i\epsilon)^{-l}$ are arbitrarily large. Consequently, this expansion
cannot be controlled. The expansion in powers of $\eta^{-1}$, on the other hand,
has always small terms for a smoothly varying ${\bf U}^{-1}$, provided $\eta$ is not too 
small in comparison with energies of these modes. 
The expansion in powers of $\eta^{-1}$ can also be justified by using large values
of $\eta$ formally and then continue it down to physically reasonable
values of $\eta$. Since no singularity appears, this extrapolation is at least qualitatively
correct.
Thus the second-order term is a good approximation for low energies. 
It gives for the expression in Eq. (\ref{det1})
\beq
S_2= -{8\over\eta}\sum_{r,r'}(\epsilon\delta_{rr'}+C_{rr'})
\psi_{r}{\bar\psi}_{r'} +o(\eta^{-3}) \ , 
\label{exp}
\eeq
where $C_{rr'}$ is
\beq
{1\over 2\eta}\Big(
\sum_{{\bar r}}{\rm Tr}_2\left[
H_{0,{\bar r}r'}H_{0,r'{\bar r}}\right]
\delta_{r,r'}-{\rm Tr}_2\left[
H_{0,rr'}H_{0,r'r}\right]
\Big) \ .
\label{realstrong}
\eeq
The average 2PGF of Eq. (\ref{green22}) then reads
\beq
K(r,r';\epsilon)
\approx {\eta^3\over 2g^2}(\epsilon + C)^{-1}_{rr'} \ .
\label{corr1}
\eeq
This is the main result of our SP calculation. It means that
details of the transport properties depend only on the SP solution $\eta$ and 
the average Hamiltonian $H_0$ in the correlation $C$. A similar expression was
found for the 2PGF of weakly disordered Dirac fermions \cite{ziegler98}.

{\it Discussion:} 1) $B=0$: Translational invariance of $C_{rr'}$ suggests to
study its Fourier components
\beq
C(q)=\frac{1}{\eta}
\left(2-\cos q_2-\cos[(\sqrt{3}q_1+q_2)/2]\right) \ , 
\label{disp2}
\eeq
or for the more general case (e.g. for a low-energy approximation of $H_0$ by
Dirac fermions), the Fourier components
\[
C(q)
=\frac{1}{2\eta}\int_k\left[{\rm Tr}_2(h_{k}^2)-{\rm Tr}_2(h_{k+q/2}h_{k-q/2})
\right] \ .
\]
Here $h_{k}$ are the Fourier components of $H_0$, i.e. for 
Dirac fermions we have $h_k=k_1\sigma_1+k_2\sigma_2$.
$C(q)$ vanishes at $q=0$ and describes the dispersion of the two-particle excitations.
In constrast to quasiparticles with Hamiltonian $H_0$, which have low-energy 
excitations near the Dirac nodes $k\ne 0$, the two-particle excitations at low energies 
appear around $q=0$. This implies that $K(q;\epsilon)$ is a diffusion propagator with 
diffusion tensor
\beq
D_{ij}=\frac{1}{2}\frac{\partial^2C(q)}{\partial q_i\partial q_j}\Big|_{q=0}=
\frac{2}{\eta}\int_k {\rm Tr}_2\left(\frac{\partial h_{k}}{\partial k_i}
\frac{\partial h_{k}}{\partial k_j}\right) \ .
\label{diffcoeff}
\eeq
Then the main contribution to diffusion comes from the vicinities of the
Dirac points, where the slope of $h_k$ is the strongest. This would justify
an approximation of $h_k$ by Dirac fermions.
Furthermore, in agreement with physical intuition,
$D$ is monotoneously decreasing with increasing disorder.
Finally, the DC conductivity $\sigma$ can be evaluated via the Kubo formula as 
\beq
\sigma_{jj}\propto 
-\epsilon^2\frac{\partial^2}{2q_j^2}K(q;\epsilon)\Big|_{q=0,\epsilon=0}
\propto \frac{\eta^2}{g^2}
\eeq
in units of $e^2/h$. This result indicates that the minimal conductivity is not constant but
decreases with increasing disorder. For the regime of strong disorder the conductivity
is proportional to $g^{-1}$, which was also found for topological disorder \cite{cortijo06}.

2) $B\ne0$: 
The Hamiltonian $H_0$ appears in $C_{rr'}$ only as
\[
{\rm Tr}_2\left[H_{0,rr'}H_{0,r'r}\right]=|d_{rr'}(B)|^2+|d_{r'r}(B)|^2 \ .
\]
For a given pair of nearest neighbors $(r,r')$ we have only one contribution for
each pair, namely 
$|e^{i\phi_{r,j}}\delta_{r',r\pm c_j}|^2$, such that the Peierls phase drops out:
$|d_{rr'}(B)|^2=|d_{rr'}(0)|^2$. 
Thus, the magnetic field dependence of the average 2PGF in Eq. (\ref{corr1}) is due
to the SP solution $\eta$. The latter can be evaluated from Eq. (\ref{spe1}) by
expanding the right-hand side in powers of $\eta^{-1}$:
\beq
\eta^2=4g-\frac{4g}{\eta^2}\sum_{r'}{\rm Tr}_2\left[H_{0,rr'}H_{0,r'r}\right]
+o(\eta^{-3})
\ .
\label{spe22b}
\eeq
Again, the $B$ dependence appears only in higher orders of the expansion.
While $\eta$ increases monotoneously with $g$, its $B$ dependence is suppressed.

Disorder due to potential scattering by impurities has been ignored in our approach. Including
it in the calculation
would affect transport properties at low temperatures significantly, since it causes
localization \cite{aleiner06}. It is not clear, however, whether or not this type of disorder 
is relevant in graphene at the Dirac point.  

We conclude that quasiparticles on the honeycomb lattice with 
random bonds are subject to diffusion. The diffusion coefficient as well as the
conductivity always decrease with increasing disorder. 
For sufficiently strong disorder the magnetic-field dependence drops out of the transport 
quantities. This reflects the suppression of an external magnetic field by disorder,
which is consistent with experimental observations of magnetotransport in graphene 
\cite{morozov06,cho07}.

\begin{acknowledgments}
I would like to thank Balazs D\'ora, Andre Geim, Mikhael Katsnelson, and Maria Vozmediano 
for stimulating discussions.
\end{acknowledgments}


\begin{thebibliography}{99}

\bibitem{novoselov05}
K.S. Novoselov {\it et al.}, Nature {\bf 438}, 197 (2005)

\bibitem{zhang05}
Y. Zhang {\it et al.}, Nature {\bf 438}, 201 (2005)

\bibitem{geim07}
A.K. Geim and K.S. Novoselov, Nature Materials, {\bf 6}, 183 (2007)

\bibitem{altshuler}
B.L. Altshuler and B.D. Simons, {\sl Physique quantique m\'esoscopique},
eds. E. Akkermans et al. (Elsevier, Amsterdam 1995)

\bibitem{morozov06}
S.V. Morozov {\it et al.}, 
Phys. Rev. Lett. {\bf 97}, 016801 (2006)

\bibitem{castroneto07b}
A.H. Castro Neto {\it et al.}, cond-mat/0709.1163

\bibitem{cho07}
S. Cho and M.S. Fuhrer, cond-mat/0705.3239

\bibitem{ludwig94}
A.W.W. Ludwig {\it et al.}, Phys. Rev. B {\bf 50}, 7526 (1994)

\bibitem{ziegler98}
K. Ziegler, Phys. Rev. Lett. {\bf 80}, 3113 (1998)

\bibitem{cortijo06}
A. Cortijo and M.A.H. Vozmediano, cond-mat/0709.2698

\bibitem{dejuan07}
F. de Juan, A. Cortijo, and M.A. Vozmediano, Phys. Rev. B {\bf 76}, 165409 (2007)

\bibitem{katsnelson07}
M. I. Katsnelson, A. K. Geim., Phil. Trans. R. Soc. A {\bf 366}, 195 (2008)

\bibitem{castroneto07}
A.H. Castro Neto and E.-A. Kim, cond-mat/0702562 

\bibitem{dora08}
B. D\'ora, K. Ziegler, and P. Thalmeier, cond-mat/0711.3748, Phys. Rev. B (in press)

\bibitem{hewson}
A.C. Hewson, {\sl The Kondo Problem to Heavy Fermions} (Cambridge Univ. Press 1993)

\bibitem{auslender07}
M. Auslender and M.I. Katsnelson, Phys. Rev. B {\bf 76}, 235425 (2007)

\bibitem{altland97}
A. Altland and M.R. Zirnbauer, Phys. Rev. B {\bf 55}, 1142 (1997)

\bibitem{garcia06}
A.M Garcia-Garcia and E. Cuevas, Phys. Rev. B {\bf 74}, 113101 (2006)

\bibitem{ziegler06}
K. Ziegler, Phys. Rev. Lett. {\bf 97}, 266802 (2006)

\bibitem{peres06}
N.M.R. Peres, F. Guinea, and A.H. Castro Neto, Phys. Rev. 
B {\bf 73}, 125411 (2006)

\bibitem{katsnelson05}
M.I. Katsnelson, 
Eur. Phys. J. B {\bf 51}, 157 (2006)

\bibitem{gusynin05}
V.P. Gusynin and S.G. Sharapov, Phys. Rev. Lett. {\bf 95}, 146801 (2005)

\bibitem{falkovsky06}
L.A. Falkovsky and A.A. Varlamov, cond-mat/0606800

\bibitem{cserti06}
J. Cserti, Phys. Rev. B {\bf 75}, 033405 (2007)

\bibitem{ziegler97}
K. Ziegler, Phys. Rev. B {\bf 55}, 10661 (1997)

\bibitem{ziegler96}
K. Ziegler, M.H. Hettler, and P.J. Hirschfeld, Phys. Rev. B {\bf 57}, 10825 (1998)

\bibitem{aleiner06}
I.L. Aleiner and K.B. Efetov, Phys. Rev. Lett. {\bf 97}, 236801 (2006); 
A. Altland, Phys. Rev. Lett. {\bf 97}, 236802 (2006);
P.M. Ostrovsky {\it et al.}, Phys. Rev. B {\bf 74}, 235443 (2006)

\end{thebibliography}
\end{document}